\documentclass[10pt, journal, letterpaper]{IEEEtran} %  \documentclass[10pt, conference, letterpaper]{ IEEEtran }
\usepackage{amsmath,amssymb,amsfonts}
\usepackage{algorithmic}
\usepackage{graphicx}
\usepackage{textcomp}
\usepackage{wrapfig}
\def\BibTeX{{\rm B\kern-.05em{\sc i\kern-.025em b}\kern-.08em
    T\kern-.1667em\lower.7ex\hbox{E}\kern-.125emX}}
\usepackage{float}
\usepackage{amsmath}
\usepackage{amssymb}
\usepackage{algorithmic}
\usepackage{graphicx}
\usepackage{textcomp}
\usepackage{xcolor}
\definecolor{subsectioncolor}{RGB}{0, 0, 255} % Defines subsectioncolor as blue
\usepackage{verbatim }
\usepackage{siunitx}

\usepackage{booktabs}
\usepackage{hyperref}
\usepackage{balance}
\usepackage{url}
\usepackage{tabularx}
\usepackage{array}
\usepackage{scalerel}
\usepackage{tikz}
\usetikzlibrary{svg.path}
\definecolor{orcidlogocol}{HTML}{A6CE39}
\tikzset{
  orcidlogo/.pic={
    \fill[orcidlogocol] svg{M256,128c0,70.7-57.3,128-128,128C57.3,256,0,198.7,0,128C0,57.3,57.3,0,128,0C198.7,0,256,57.3,256,128z};
    \fill[white] svg{M86.3,186.2H70.9V79.1h15.4v48.4V186.2z}
                 svg{M108.9,79.1h41.6c39.6,0,57,28.3,57,53.6c0,27.5-21.5,53.6-56.8,53.6h-41.8V79.1z M124.3,172.4h24.5c34.9,0,42.9-26.5,42.9-39.7c0-21.5-13.7-39.7-43.7-39.7h-23.7V172.4z}
                 svg{M88.7,56.8c0,5.5-4.5,10.1-10.1,10.1c-5.6,0-10.1-4.6-10.1-10.1c0-5.6,4.5-10.1,10.1-10.1C84.2,46.7,88.7,51.3,88.7,56.8z};
  }
}
\newcommand\orcidicon[1]{\href{https://orcid.org/#1}{\mbox{\scalerel*{
\begin{tikzpicture}[yscale=-1,transform shape]
\pic{orcidlogo};
\end{tikzpicture}
}{|}}}}

\usepackage{svg}

\begin{document}
\title{
Prediction of Received Power in Low-Power and Lossy Networks Deployed in  Rough Environments 
}

\author{Waltenegus Dargie\orcidicon{0000-0002-7911-8081}, \IEEEmembership{Senior Member, IEEE} 
    \thanks{Manuscript resubmitted on 06 December 2024.}
    %\thanks{This work has been partially funded by the German Research Foundation (DFG) under project agreements DA 1211/7-1.}
    \thanks{W. Dargie is with the Faculty of Computer Science, Technische Universit{\"a}t Dresden, 01062 Dresden, Germany (e-mail: waltenegus.dargie@tu-dresden.de)}
    %\thanks{C. Poellabauer is with the Knight Foundation School of Computing and Information Sciences at Florida International University, USA, (e-mail:  cpoellab@fiu.edu)}
}
\maketitle

\begin{abstract}
    Cost-efficient and low-power sensing nodes enable to monitor various physical environments. Some of these impose extreme operating conditions, subjecting the nodes to excessive heat or rainfall or motion. Rough operating conditions affect the stability of the wireless links the nodes establish and cause a significant amount of packet loss. Adaptive transmission power control (ATPC) enables nodes to adapt to extreme conditions and maintain stable wireless links and often rely on knowledge of the received power as a closed-feedback system to adjust the power of outgoing packets. However, in the presence of a significant packet loss, this knowledge may not reflect the current state of the receiver. In this paper we propose a lightweight n-step predictor which enables transmitters to adapt transmission power in the presence of lost packets. Through extensive practical deployments and testing we demonstrate that the predictor avoids expensive computation and still achieves an average prediction accuracy exceeding 90\% with a low-power radio that supports a transmission rate of 250 kbps (CC2538) and 85\% with a low-power radio that supports 50 kbps (CC1200).  
\end{abstract}


\begin{IEEEkeywords}
Adaptation, low-power networks, link quality prediction, received power, Internet-of-Things 
\end{IEEEkeywords}


\section{Introduction}
\label{intro}

Deploying low-power IoT sensing nodes in different physical environments enables to monitor vital parameters without the need for human presence or interference \cite{dargie2010fundamentals, li2020electromagnetic}. %In industry, they can monitor chemical leaks and the movements of objects \cite{shu2016toxic, dargie2023monitoring}. In smart environments, they can sense various physical parameters pertaining to air and water quality as well as pollution and flooding \cite{jamil2015smart, adamo2014smart}. 
Some of these environments impose extreme operating conditions, affecting the performance and the lifetime of the nodes \cite{wang2020active}. For example, during water quality monitoring, some nodes have to be deployed on the surfaces of restless waters, which constantly move and displace the nodes. Besides affecting the quality of the wireless links the nodes establish, the constant movement of the nodes also steadily modifies the topology of the network, making the discovery of new routes and the maintenance of existing routes a challenging assignment. One of the most important requirements for low-power sensing nodes to operate in these types of environments is dynamic adaptation of transmission power. 

A transmitter should have some knowledge of the relative distance of the receiver and the transmission path to estimate the power with which outgoing packets should be transmitted. Assuming the existence of a symmetric, unicast channel between the transmitter and the receiver, the former can estimate the relative distance of the later from the received power of ACK packets. The only problem is that, since the wireless channel is lossy, some packets will inevitably be lost and the receiver may have changed its location since its last successful transmission of ACK packets. In general, statistical-based predictors can be used to estimate the received power in the presence of lost packets, but they are computationally expensive. For example, predictors based on Minimum Mean Square Estimation (MMSE)  require matrix inversion to determine model parameters, but the computational resources (CPU, memory) this requires is something most existing resource-constrained sensing devices cannot meet. It is, therefore, important to use a predictor whose (1) computational cost is modest but (2) performance is high. The aim of this paper is to achieve these goals.
\begin{itemize}
\item As the first contribution of this paper, our predictor avoids matrix inversion through (a) function approximation and (b) orthogonalization. 
\item As the second contribution, the model tolerates multiple successively lost packets, achieving a prediction accuracy exceeding 90\% even when the packet transmission rate is modest (10 packets per second). 
\item As a third contribution, we demonstrate the usefulness of the predictor through extensive experiments involving actual deployments on four different water bodies.
\end{itemize} %, namely, a small lake on the main campus of Florida International University (FIU), North Biscayne Bay, South Florida, Miami South Beach, and Miami Crandon Beach. 

The remaining part of the paper is organized as follows. In Section \ref{sec:related}, we review related work. In Section \ref{sec:deployment}, we describe the deployment scenarios. In Section~\ref{sec:background} we briefly discuss the autocorrelation function, as our model relies on it. In Section~\ref{sec:model}, we present our model, define the model parameters and discuss different approaches to minimise the computation cost of the model. In Section~\ref{sec:evaluation} we evaluate the performance of our model and compare its performance with some competitive work. Finally in Section~\ref{sec:conclusion} we provide concluding remarks and outline future work.

%In a recent study, we experimentally demonstrated that the change in the received power of low-power sensing nodes deployed on the surface of different water bodies was proportional to the motion of the underlying water. Moreover, we proposed a Minimum Mean Square Estimation which takes statistics of past received power and 3D accelerations to predict the received power. The model consistently achieves a prediction accuracy exceeding 97\%. Nevertheless, it has the following shortcomings: 
%\begin{itemize}
%    \item A high communication and processing cost due to the model's reliance on 3D inertial statistics.
 %   \item The need to synchronise time between the receiver and the transmitter, as the model relies on linear correlation between the received power and the 3D acceleration.
 %   \item A high computation cost involving matrix inversion.
%\end{itemize}
%These shortcomings make long-term adaptation challenging. Operating nodes without a dynamic transmission power adaptation is equally disadvantageous because a considerable amount of packets will be lost during communication. 
\section{Related Work}
\label{sec:related}

Managing the power consumption of low-power sensing devices is of paramount importance to achieve reliable communication and long operation life \cite{Trihinas8057144, azarhava2020energy}. In most existing architectures, the radio subsystem of these devices is second only to the processing subsystem when it comes to power consumption \cite{dargie2011dynamic}. In the literature, the problem of power consumption is addressed in different ways, including, micro energy harvesting \cite{Ju7842599}, energy prediction and task scheduling \cite{liu2020multistep, Arghavani10404000}, compressed sensing \cite{Zhao10577633}, wireless power transfer \cite{clerckx2021wireless}, topology control \cite{shahraki2020clustering}, and adaptive duty cycles \cite{amutha2020wsn}. Deployment environments (such as human presence and activities) and weather conditions (excessive heat and rainfall) affect the performance of low-power sensing networks, causing serious link quality fluctuations, packet loss, and end-to-end packet transmission latency \cite{dargie2024estimation, hu2020tngan, zonzini2020vibration}. 

%Similar to the present work, the work in~\cite{10.5194/acp-9-2413-2009} and \cite{10.1145/1089444.1089466} investigates the impact of weather and atmospheric conditions on the quality of wireless links. The work in \cite{10.18280/mmep.080316} investigates the performances of low-power wireless devices deployed near water surfaces and attempts to quantify the impact of water on signal absorption and scattering. In~\cite{9706046}, the authors investigate the impact of recurrent natural phenomena (tides and waves) on the quality of wireless links and how the motion of water affects signal propagation and interference patterns. Gaitán et al.~\cite{9894283} investigate large-scale fading dynamics of LoRa line-of-sight (LoS) links deployed over an estuary having characteristic intertidal zones. The authors consider both shore-to-shore and shore-to-vessel communications. The proposed model predicts path-loss by taking into consideration spatial and temporal as well as physical aspects of the RF signal's interaction with environmental dynamics.

In \cite{park2020adaptable} a dynamic rate and transmission power control algorithm for Bluetooth Low Energy (BLE) is proposed. Based on the latency and throughput requirements of a client device, the protocol aims to adjust the connection interval, data rate, and transmission power according to the link state. The protocol classifies a wireless link into good, fair, and bad. In the first, the requirements of the client can be fulfilled and there is room either to increase transmission rate or decrease transmission power. In the second, the link quality is just enough to fulfil the requirements and there is no need to make any adaptation. In the third, the perceived quality of the link is not sufficient to satisfy the latency requirement of the client application; this requires either a reduction of the transmission rate or an increase of the transmission power. In either case, rate and transmission power adaptation take place gradually, until the desired link state is achieved, thus eliminating the need to directly estimate/predict the received power. However, the gradual increase/decrease of the transmission power or the transmission rate takes time and may cause a considerable packet loss in case the wireless link experiences drastic changes.

In \cite{cao2020optimized}, a model based on MMSE is proposed to jointly optimize the transmit power of multiple low-power devices (the transmitters) and the signal scaling (denoising) factor of a fusion centre (FC), subject to individual average power constraints at the devices. The proposed model works for a single-hop, star-topology network wherein the FC plays the role of an aggregator whose purpose is to fuse the information from the low-power devices. The topology enables the devices to simultaneously transmit, however, instead of transmitting the original signal, the devices choose a function which normalises the original information, such that the aggregate information at the FC has a normally distributed random process having a zero mean and a variance of one. This configuration enables the devices to reduce the cost of transmission and, the FC, to reconstruct the original information in the presence of signal distortion. Multiple assumptions as regards the channel -- time-variant vs. time-invariant; with or without the availability of channel state information (CSI) -- and the number of simultaneously transmitting devices lead to multiple optimisation solutions, some of which are rather computationally expensive. With simulation results, the authors demonstrate that multiple trade-offs between model complexity, transmission power, and aggregation accuracy (in terms of mean square error) can be achieved.  

 In \cite{Arghavani10404000}, a Gilbert-Elliott Markov chain model is employed to monitor channel fluctuations in low-power wearable networks and to predict long-term channel states. Based on a predicted state, packets are either transmitted with the lowest transmission power possible or differed and locally buffered without violating a set deadline. In \cite{zang2017gait}, the transmission power of a wearable network is dynamically adapted according to the underlying gait pattern. The pattern (periodicity) is learned from the linear acceleration of the motion of the user and correlated with the RSSI values of incoming packets. This enables to estimate the time offset between the acceleration peak and the corresponding RSSI peak. Then, transmissions are scheduled at the channel peaks to achieve a high packet delivery ratio (PDR) using the lowest transmission power possible.

A work closer to ours is the one proposed by Lin et al. \cite{lin2016atpc}, in which the authors  empirically demonstrate that the quality of low-power wireless links considerably varies even for static deployments and that previous topology control solutions based on fixed transmission power budgets are inadequate to achieve reliable and efficient communication. To address this concern, the authors propose a transmission power control scheme which adapts transmission power to environmental dynamics. Accordingly, each node in a wireless sensor network builds a link quality model for each of its neighbours, linearly correlating transmission power with two received power metrics (RSSI and Link Quality Indicator) using least square approximation. The authors report an impressive performance which reduced overall power consumption by up to 53.6\% compared to solutions based on {\em maximum transmission policy} and a 99\% of packet Reception Ratio (PRR). The evaluation consisted of deployments carried out on a grassy meadow, a parking lot, and a corridor.

\section{Deployment}
\label{sec:deployment}

\begin{figure}
	\centering
	\includegraphics[width=0.45\textwidth]{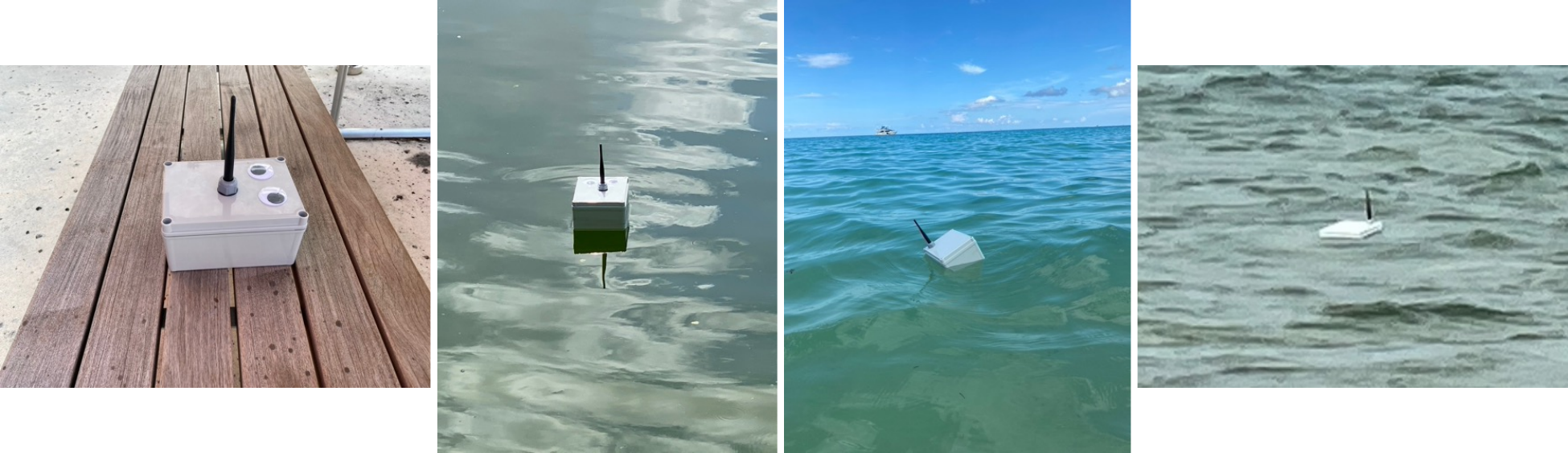}
	\caption{Low-power and waterproof IoT sensing nodes deployed on the surface of different water bodies.}
	\label{fig:deployment}
\end{figure}

In order to experimentally investigate how wireless link quality fluctuates in harsh and extreme environments, we deployed low-power wireless sensor networks on the surface of different water bodies in Miami, Florida---a small lake on the main campus of Florida International University (FIU), North Biscayne Bay, Crandon Beach, and Miami South Beach. We placed the nodes inside waterproof boxes and installed waterproof marine antennas, so that the nodes can communicate in harsh weather and rough operating conditions (ref. to Fig.~\ref{fig:deployment}). Each sensor platform  integrates two different types of radios. One of them, the CC1200,\footnote{\url{https://www.ti.com/product/de-de/CC1200} Last visit. November 30, 2024, 06:26 PM, CET.} can be configured to  operate in different sub-Gigahertz frequency bands (169, 433, 868, 915, and 920 MHz) and is capable of data buffering, burst transmissions, clear channel assessment, and Wake-On-Radio. In all our experiments, the radio was configured to operate in the 869.5 MHz band (the so-called low-power mode). The maximum transmission power in this band is 16 dBm. The radio's sensitivity depends on its transmission rate: –123 dBm at 1.2 kbps and –109 dBm at 50 kbps. According to the specification, the CC1200 has a maximum transmission range of 4 km, our experience suggests, however, that the practically achievable range is much less than this value ($\leq$ 1 km), depending on both environmental factors and network configuration. Similarly, the CC2538 system-on-chip (SoC) radio\footnote{\url{https://www.ti.com/product/CC2538} Last visit. November 30, 2024, 06:42 PM, CET.} integrates a 2.4 GHz IEEE 802.15.4-compliant RF transceiver with a sensitivity of –97 dBm and an adjustable output power (max. output power = 7 dBm). The radio transmits at 250 kbps nominal rate. Compared to the CC1200 radio, it is much more stable. However, for most practical purposes, the achievable transmission range is less than 100 m in free space.  For both radios, the packet size was 128 bytes. With this packet size the sustainable rate that was supported by the CC1200 was 2 packets per second; for the CC2538, it was 10 packets per second.

\begin{figure}
	\centering
	\includegraphics[width=0.45\textwidth]{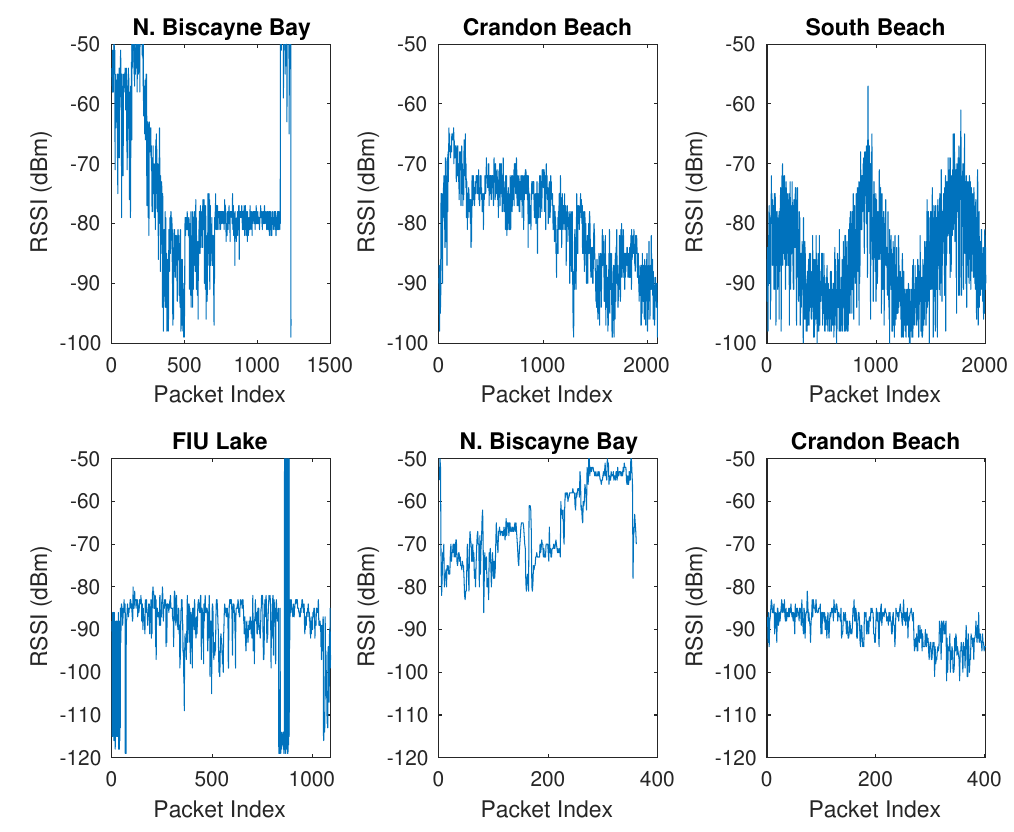}
	\caption{Link quality fluctuation due to the motion of water. TOP: CC2538 SoC. BOTTOM: CC1200 Radio.}
	\label{fig:lqf}
\end{figure}

In each location, the nodes self-organised to establish a multi-hop wireless sensor network. In each network there were five floating sensor nodes and an additional static node outside the water, serving as a gateway node. Due to the constant and significant motion of the underlying water surface, the quality of the wireless links the nodes established changed considerably, leading to frequent disconnections and a considerable amount of packet loss (more than 30\%). The different water bodies affected the wireless links differently, mainly due to their difference in motion. The lake was relatively calm but three artificial fountains in its midst created continuous circular ripples, thereby locally oscillating the sensor nodes. The surface of the water in North Biscayne Bay was moved by modest waves the direction of which was frequently disturbed by large boats and yachts driving nearby. The waters of Crandon Beach and Miami South Beach were, by comparison, rough. The waves at Crandon Beach were short and rapid; the waves at South Beach were long and considerably large. Fig.~\ref{fig:lqf} displays link quality fluctuation (the change in the RSSI values of received packets) for the different deployment settings and radios.

\begin{figure}
    \centering
    \includegraphics[width=0.45\textwidth]{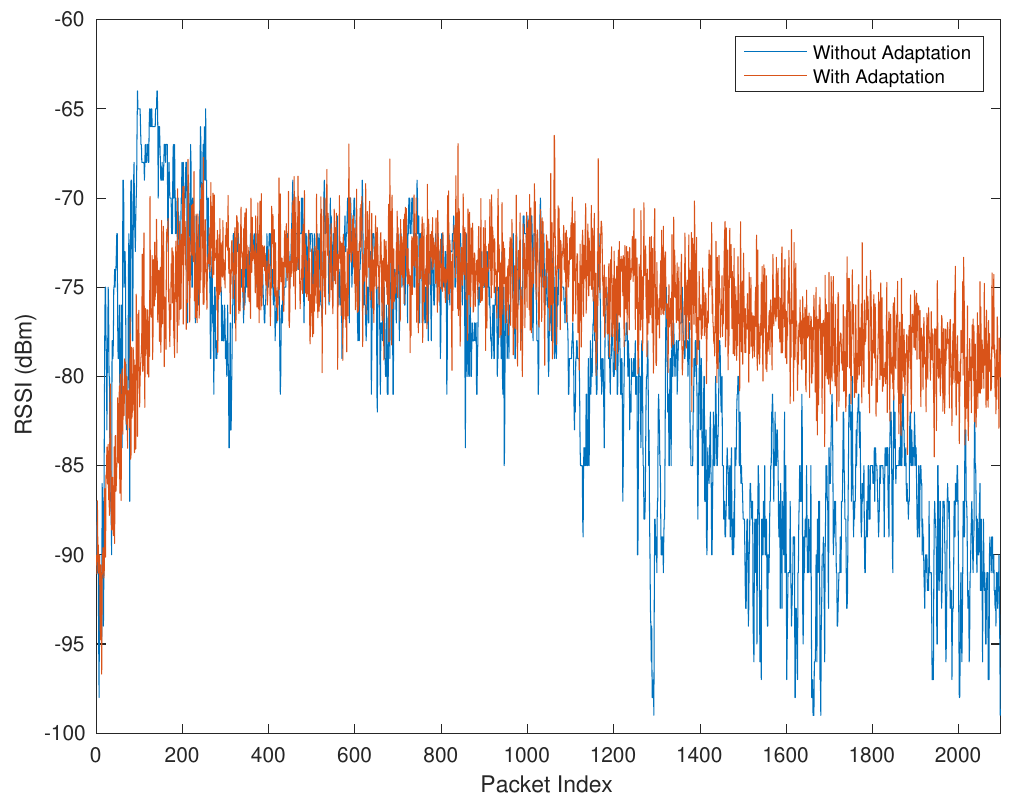}
    \caption{Comparison of two types of received powers. The blue line indicates a strong fluctuation in the absence of an adaptive transmission power. The red line describes a scenario in which the transmission power is adapted to the underlying condition, so that the received power is always above a set threshold.}
    \label{fig:dynamic_adaptation}
\end{figure}

Fig.~\ref{fig:dynamic_adaptation} illustrates what we aim to achieve. The plot with the blue line indicates a received power in the absence of adaption of the transmission power. Both the transmitter and the receiver nodes were deployed on the surface of the Atlantic Ocean at Crandon Beach, Miami, Florida, and constantly swayed and carried by the water waves. The plot with the red lines indicates the received power when an adaptive transmission power was in place. As can be seen, the transmitter adjusted its power, so that packets could be received by the transmitter with a power the magnitude of which was above a set threshold. 
\section{Background}
\label{sec:background}

The received power is affected by different dynamic physical factors and should be regarded as a stochastic process, $\mathbf{r}(t)$. Its predictability can be determined by its autocorrelation, assuming that it can be taken as a wide-sense stationary stochastic process (WSS):

\begin{align}
    \label{eq:b1}
    R_{rr} \left ( t_1, t_2 \right ) & = E \left \{ \mathbf{r}(t_1) \mathbf{r}(t_2) \right \} \\ \nonumber
                                & = \int_{-\infty}^{\infty} \int_{-\infty}^{\infty} r_1, r_2 f \left (r_1, r_2; t_1, t_2 \right ) dr_1 dr_2
\end{align}
where the bold-face letters $\mathbf{r}(t_1)$ and  $\mathbf{r}(t_2)$ refer to the received power at times $t_1$ and $t_2$, respectively; the plane letters $r_1$ and $r_2$ are arbitrary real values the random variables take; and $f$ is the joint probability density function of $\mathbf{r}(t_1)$ and $\mathbf{r}(t_2)$. For a WSS process, the autocorrelation function is insensitive of time-shifts, depending only on the difference between $t_1$ and $t_2$: $\tau = t_2 - t_1$:

\begin{equation}
    \label{eq:b2}
    R_{rr} \left ( \tau \right ) = E \left \{ \mathbf{r}(t) \mathbf{r}(t + \tau ) \right \}
\end{equation}
From Equations~\ref{eq:b1} and \ref{eq:b2}, it can be seen that the autocorrelation is an expected value.  

\begin{figure}
	\centering
	\includegraphics[width=8cm,height=9.5cm, angle = 270]{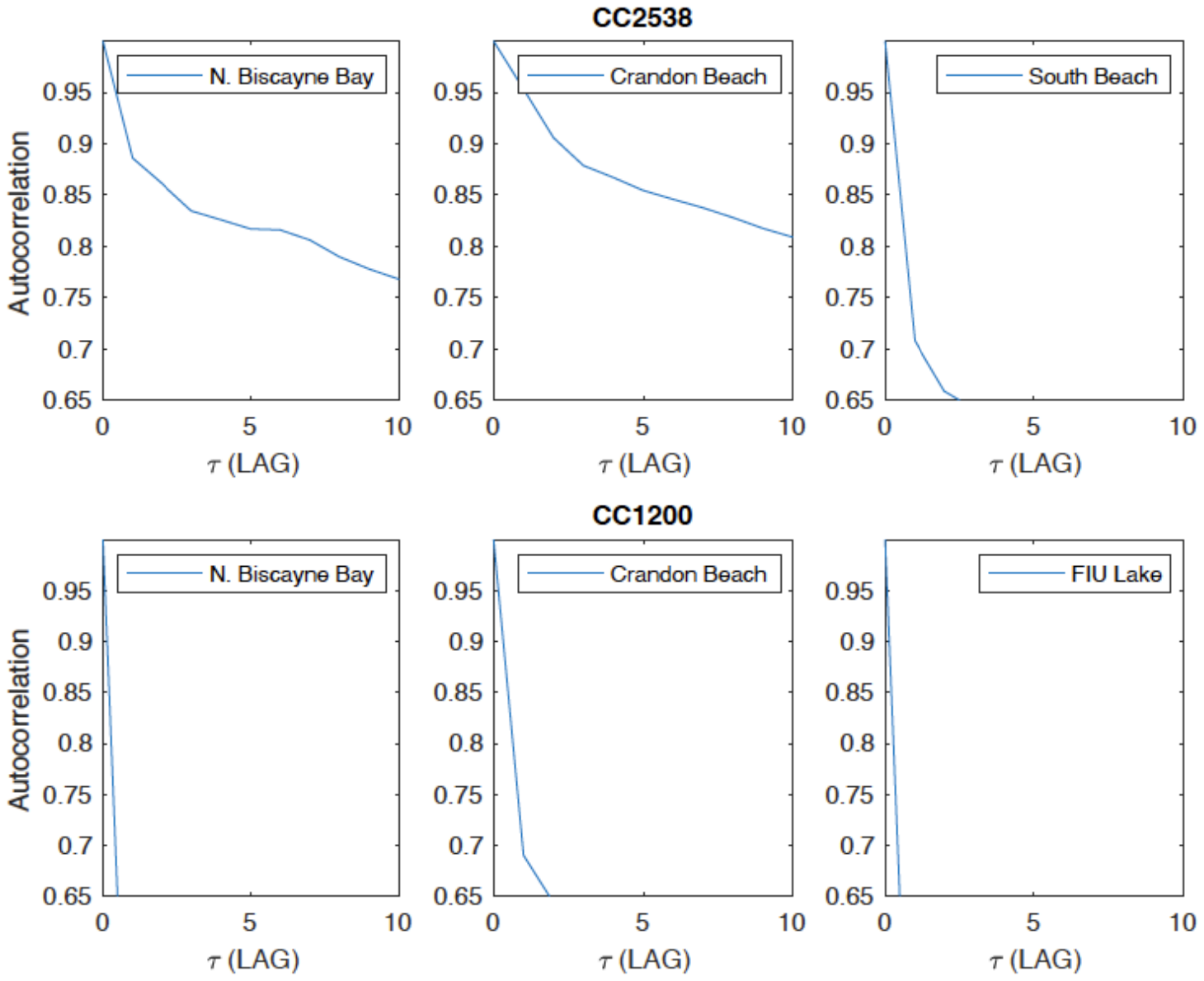}
	
	\caption{The autocorrelation function of the received power for the different deployment environments. TOP: CC1200. BOTTOM: CC2538.}
	\label{fig:acf}
\end{figure}

Fig.~\ref{fig:acf} shows the one-side ($\tau \geq 0$) autocorrelation functions of the received power of different wireless links for the two radios we employed in our experiments. In each case, 2000 packets were transmitted to establish the statistics. The plots suggest that the wireless links established with the CC2538 SoC exhibited strong autocorrelation. This is in part due to the relatively high sampling rate (10 Hz) with which packets were transmitted. By contrast, the CC1200 radio chip could support a sustainable rate of 2 packets per second (for a packet size of 128 bytes). In both cases, the plots suggest that for a small time lag, the autocorrelation can be employed to predict the future received power.  
\section{Model}
\label{sec:model}

The research issue we address is depicted in Fig.~\ref{fig:scenario}. A receiver sends acknowledgment packets to a transmitter. The transmitter estimates the relative distance of the receiver by evaluating the received power of these packets. Due to the nature of the wireless link, however, some of the acknowledgment packets may be corrupted or lost, in which case, the transmitter has to predict the received power using an n-step predictor. Thus, the predictor expresses the received power at time {\em t +} $\tau$, $\tau > 0$, in terms of the received power at time $t$ and the statistics of the change the transmission power experienced immediately before that. Our model assumes that compared to the rate at which packets are transmitted the change the physical environment imposes on the deployed nodes (for example, the physical movement of the nodes), is slower. Hence, for a small $\tau$, we can express the change in the received power at time $t + \tau$  in terms of the received power at time $t$ and its derivation at $t$:

\begin{figure}[t!]
	\centering
	\includegraphics[width=0.4\textwidth]{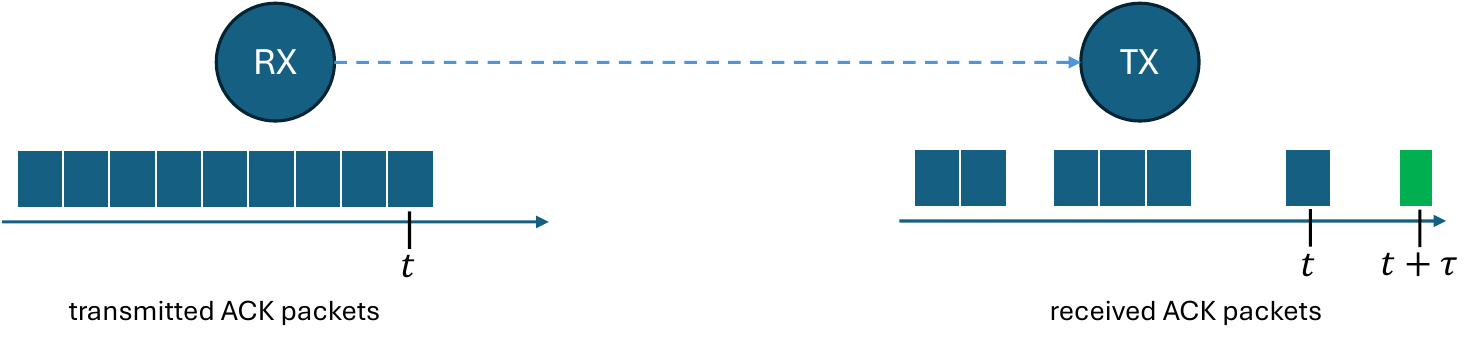}
	\caption{Communication scenario.}
	\label{fig:scenario}
\end{figure}

\begin{equation}
    \label{eq:m1}
    \hat{ \mathbf{r} } \left ( t + \tau \right ) = \rho_{r} \mathbf{r} \left ( t \right ) + \rho_{r'} \mathbf{r}' \left ( t \right )
\end{equation}
For a normal variable, Equation~\ref{eq:m1} resembles a Taylor's Approximation. However, for our case, both $\mathbf{r}$ and $\mathbf{r}'$ are known to us only in a probabilistic sense. Hence, the determination of  $\hat{\mathbf{r}} (t + \tau )$ must consist of the joint statistics of the two random variables.   

\begin{table}[hbt!]
        \caption{Summary of the variables used by the model.}
    \label{tab:var}
    \centering
    \begin{tabular}{|l|l|}
    \hline
        Variable & Explanation \\
        \hline
        \hline
        $\mathbf{r}(t)$                 & Received power at $t$ (in dBm)               \\
        $\hat{\mathbf{r}}(t)$           & The best estimate of $\mathbf{r}(t)$      \\
        $\mathbf{e}(t)$                 & $\mathbf{r}(t) - \hat{\mathbf{r}}(t)$\\
        $\mathbf{r}' (t)$               & $\frac{d  \mathbf{r}(t) }{dt}$      \\
        $\rho_r$                        & The weight assigned to  $\mathbf{r}(t)$      \\
        $\rho_{r'}$                     & The weight assigned to  $\mathbf{r}'(t)$      \\
        $\mathbf{R}_{rr}(t_1, t_2)$      & $E \left \{ \mathbf{r}(t_1) \mathbf{r}(t_2) \right \}$ \\
        $\mathbf{R}_{rr}'(\tau$)         & $\frac{d  \mathbf{R}_{rr}(\tau)}{d \tau}  $    \\
        $\mathbf{R}_{rr'}(t_1, t_2)$     & $E \left \{ \mathbf{r}(t_1) \mathbf{r}'(t_2) \right \}$ \\
        $\mathbf{R}_{rr}''(\tau)$        & $\frac{d \mathbf{R}_{rr}(\tau)}{d\tau^2} $  \\
        $\mathbf{R}_{r'r'}(\tau)$        & $E \left \{ \mathbf{r}'(t + \tau) \mathbf{r}'(t) \right \}$ \\
        \hline
    \end{tabular}
\end{table}

\subsection{Application of MS Estimation}
In order to determine the coefficients $\rho_r$ and $\rho_{r'}$, we propose Minimum Mean Square Estimation (the variables and parameters we require for our model are summarised in Table~\ref{tab:var}). Thus, the mean square error is given as:
\begin{equation} \label{eq:m2}
E \left \{ \mathbf{e}^2(t + \tau) \right \} = E \left \{ \left [  \mathbf{r} \left (t + \tau \right )  -  \hat{ \mathbf{r}} \left (t + \tau \right ) \right ]^2 \right \} 
\end{equation}
The coefficients which minimize the mean square error are determined by (1) differentiating Equation~\ref{eq:m2} with respect to $\rho_{r}$ and $\rho_{r'}$ and (2) setting the results to zero:
\begin{align}
    \label{eq:m2b}
    \frac{\partial}{\partial \rho_r} E \left \{ \mathbf{e}^2(t + \tau) \right \} = E \left \{ \mathbf{e}(t  + \tau) \mathbf{r} (t) \right \} = 0
    %\frac{\partial}{\partial \rho_{r'}} E \left \{ \mathbf{e}^2(t) \right \} = 0
\end{align}
The pattern is the same for $\rho_{r'}$. Notice that the error is orthogonal to the data. This is known as the {\em orthogonality principle} \cite{zhang2022modern, papoulis2002probability}. Thus, we have two equations for the two unknown coefficients, leading to the following expression:
\begin{equation} 
\label{eq:m3}
\begin{bmatrix}
    R_{rr} (\tau) \\
    R_{rr'} (\tau) \\
\end{bmatrix} = 
\begin{bmatrix}
     R_{rr} (0) &  R_{rr'} (0)  \\
     R_{rr'} (0) & R_{r'r'} (0) 
\end{bmatrix}
\begin{bmatrix}
    \rho_{r}  \\
    \rho_{r'} \\
\end{bmatrix}
\end{equation}
where $R_{rr}(\tau) = E\{ \mathbf{r}(t + \tau) \mathbf{r}(t) \}$ is the correlation between the future received power and the presently received power. Similarly,  $R_{rr'}(\tau) = E\{\mathbf{r}(t + \tau) \mathbf{r}'(t) \}$ is the correlation between the future received power and the change in the received power at time $t$; $R_{rr}(0) = E\{ \mathbf{r}^2(t) \}$ is the autocorrelation of the received power; and, finally, $R_{r'r'}(0) = E\{ \mathbf{r}'(t) \mathbf{r}'(t)\}$ is the autocorrelation of the change in the received power. The matrix in Equation~\ref{eq:m3} can be established from observation alone. Taking its inverse to the term on the left sides enables to determine the optimal coefficients $\rho_r$ and $\rho_{r'}$ which are needed in Equation~\ref{eq:m1} to predict the future received power. 

\begin{figure}
	\centering
	\includegraphics[width=0.4\textwidth]{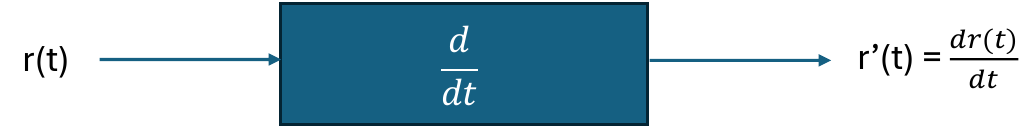}
	\caption{A differentiator system modelled as a linear time-invariant system.}
	\label{fig:ltis}
\end{figure}

Since $\mathbf{r}' (t)$ is a dependent random variable, $R_{rr'}$ and $R_{r'r'}$ can be expressed in terms of $R_{rr} (t)$. To establish the mathematical expressions, we can conceive of a linear system which takes $\mathbf{r}(t)$ as its input and produces $\mathbf{r}'(t)$, as shown in Fig.~\ref{fig:ltis}. Hence:

\begin{footnotesize}
\begin{align}
   \label{eq:m4}
    R_{rr'} \left ( t_1, t_2 \right ) = E \left \{\mathbf{r}(t_1) \mathbf{r}'(t_2) \right \} = E \left \{ \mathbf{r} (t_1) \frac{\partial \mathbf{r} (t_2) }  {\partial t_2} \ \right \}
\end{align}
\end{footnotesize}

Because of the linearity assumption, the above expression can be written as:
\begin{align}
            R_{rr'} \left ( t_1, t_2 \right ) = \frac{\partial} {\partial t_2}  E \left \{ \mathbf{r} (t_1) \mathbf{r} (t_2) \right \} = \mathbf{R}'_{rr} \left (t_1, t_2 \right ) 
\end{align}
In other words, the correlation between $\mathbf{r}(t_1)$ and $\mathbf{r}'(t_2)$ is the same as the differentiation of $R(t_1, t_2)$ with respect to $t_2$. As we have already mentioned, for a wide-sense stationary process, the autocorrelation depends only on the difference of $t_1$ and $t_2$: $\tau = t_2 - t_1$. Substituting $\tau$ in Equation~\ref{eq:m4} as follows:
\begin{align}
t_2 - t_1 = \tau\\ \nonumber
dt_2 =  d\tau \nonumber
\end{align}
yields:
\begin{align}
   \label{eq:m5}
    R_{rr'} \left (\tau \right ) & = R'_{rr} \left (\tau \right ) 
\end{align}
Similarly, 
\begin{align}
   \label{eq:m6}
    R_{r' r'} \left (\tau \right ) & = E \left \{ \mathbf{r}'(t_1 + \tau) \mathbf{r}'(t_1) \right \} \\ \nonumber
        & = \frac{d^2}{ d\tau^2 } E \left \{ \mathbf{r} (t_1) \mathbf{r} (t_1 + \tau) \right \} = - R''_{rr} \left (\tau \right ) 
\end{align}
From Equations~\ref{eq:m5} and ~\ref{eq:m6}, it can be concluded that the matrix in Equation~\ref{eq:m3} can be determined from the statistics of $\mathbf{r}$ alone. The mean square error we introduce as a result of applying Equation~\ref{eq:m1} to predict the future received power is given as:
\begin{footnotesize}
 \begin{align}    
\label{eq:m7a}
E \left \{ \mathbf{e}^2 (t  + \tau ) \right \}  & = E \left \{ \left [ \mathbf{r}(t + \tau) -   \hat{ \mathbf{r}}(t + \tau) \right ] \left [ \mathbf{r}(t + \tau) -   \hat{ \mathbf{r}}(t + \tau) \right ]  \right \}   \nonumber \\
                                   & = E \left \{ \left [ \mathbf{r}(t + \tau) -   \hat{ \mathbf{r}}(t + \tau) \right ] \mathbf{r}(t + \tau)  \right \}  -     \nonumber \\  
                                   & E \left \{ \left [ \mathbf{r}(t + \tau) - \hat{ \mathbf{r}}(t + \tau) \right ] \hat{ \mathbf{r}}(t + \tau)  \right \} 
\end{align}
\end{footnotesize}

\noindent Expanding the last term in Equation~\ref{eq:m7a} we get: 

\begin{footnotesize}
\begin{align} 
\label{eq:m7b}
E \left \{ \left [ \mathbf{r}(t + \tau) - \hat{ \mathbf{r}}(t + \tau) \right ] \hat{ \mathbf{r}}(t + \tau)  \right \} =  E \left \{ \mathbf{e}(t + \tau) \hat{ \mathbf{r}}(t + \tau) \right \}
\nonumber \\
= \rho_r E \left \{ \mathbf{e}(t + \tau)  \mathbf{r}(t) \right \} +  
 \rho_{r'} E \left \{ \mathbf{e}(t + \tau) \mathbf{r}'(t) \right \} 
\end{align}
\end{footnotesize}

\noindent From Equation~\ref{eq:m2b} it follows that the mean square error is orthogonal to both $\mathbf{r}(t)$ and $\mathbf{r}'(t)$ and, hence, Equation~\ref{eq:m7b} is zero. Consequently, the mean square error is given as:

\begin{footnotesize}
 \begin{align}    
\label{eq:m7}
    E \left \{ \mathbf{e}^2 (t + \tau ) \right \} & = E \left \{ \left [ \mathbf{r}(t + \tau) - \rho_r \mathbf{r} (t) - \rho_{r'} \mathbf{r}' (t) \right ] \mathbf{r} (t + \tau ) \right \} \nonumber \\ 
        & = R_{rr}(0) -  \rho_{r}R_{rr} (\tau) + \rho_{r'} R'_{rr}(\tau)
\end{align}
\end{footnotesize}

\subsection{Orthonormal Model Parameters}

Determining the model parameters using Equation~\ref{eq:m3} entails matrix inversion, which we wish to avoid. One way to avoid matrix inversion is to linearly transform $\mathbf{r}(t)$ and $\mathbf{r}'(t)$ in such a way that the transformed parameters are orthogonal to each other and their covariance is zero (so-called the Gram-Schmidt approach \cite{leon2013gram, papoulis2002probability}). Suppose:
\begin{align} \label{eq:m9}
\mathbf{p}_1(t) & = \rho_{11} \mathbf{r}(t) \\ \nonumber
\mathbf{p}_2(t) & = \rho_{21} \mathbf{r}(t)  + \rho_{22} \mathbf{r}'(t) 
\end{align}
under the condition:
\begin{align}\label{eq:m9b}
E \left [ \mathbf{p}_1(t) \mathbf{p}_2(t) \right ] = 0 
\end{align}
Moreover, we require,
\begin{align} \label{eq:m9c}
E \left [ \mathbf{p}^2_1(t) \right ] = E \left [ \mathbf{p}_2^2(t) \right ] = 1
\end{align}
Solving for $\rho_{11}$ is straightforward, since $E[\mathbf{p}_1^2(t) ] = \rho^2_{11} R_{rr}(0)$, which leads to:
\begin{align} \label{eq:m10}
\rho_{11} = \frac{1}{\sqrt{R_{rr}(0)} }   
\end{align}
The second expression in Equation~\ref{eq:m9} can be solved likewise, since we have two unknowns and two equations. In Equation \ref{eq:m9b}, We conditioned $\mathbf{p}_1(t)$ and $\mathbf{p}_2(t)$ to be orthogonal. Combining this fact with Equation \ref{eq:m9}, we have: 
\begin{align} \label{eq:m11}
    E \left [ \mathbf{p}_1(t) \mathbf{p}_2(t) \right ] & = \rho_{11} E \left [ \mathbf{r}(t) \mathbf{p}_2(t) \right ] = 0  \nonumber \\ 
     &  = \rho_{21}  E \left [ \mathbf{r}(t) \mathbf{r}(t)  \right ]  + \rho_{22} E \left [ \mathbf{r}(t) \mathbf{r}'(t) \right ] = 0 \nonumber \\ 
     & = \rho_{21} R_{rr}(0) - \rho_{22} R_{rr}' (0) = 0
\end{align}
Restructuring terms in Equation~ \ref{eq:m11}, we get:
\begin{align} \label{eq:m11b}
    \rho_{21} = \rho_{22} \frac{R_{rr'}(0)}{R_{rr}(0)} \\ \nonumber 
\end{align}
Moreover (from Equation~\ref{eq:m9b}), 
\begin{align}\label{eq:m12}
E \left \{ \left [ \rho_{21} \mathbf{r}(t) + \rho_{22} \mathbf{r}'(t) \right ]  \left [ \rho_{21} \mathbf{r}(t) + \rho_{22} \mathbf{r}'(t) \right ] \right \} = 1
\end{align} 
Combining Equation~\ref{eq:m11} with \ref{eq:m12} yields:
\begin{align} \label{eq:m13}
        \rho_{22} = \sqrt{\frac{R_{rr}(0)R_{r'r'}(0) - R^2_{rr'}(0)}{R_{rr}(0)}}
\end{align}
Having determined $\rho_{11}$, $\rho_{21}$, and $\rho_{22}$, we can now predict the future received power in terms of $\mathbf{p}_1(t)$ and $\mathbf{p}_2(t)$, instead of $\mathbf{r}(t)$  and $\mathbf{r}'(t)$: 
\begin{align} \label{eq:m14}
    \hat{\mathbf{r}} \left( t + \tau \right ) = \pi_1 \mathbf{p}_1(t) + \pi_2 \mathbf{p}_2(t)
\end{align}
Differentiating the mean square error due to Equation~\ref{eq:m14} in the same way we differentiated Equation~\ref{eq:m2b} results in the determination of the optimal coefficients:
\begin{align} \label{eq:m15}
    \pi_1 = \frac{R_{rr}(\tau) }{R_{rr}(0)}
\end{align}
Likewise,
\begin{align} \label{eq:m16}
    \pi_2 = \frac{\rho_{21}R_{rr}(\tau) + \rho_{22} R_{rr'} (\tau ) }{\rho_{21}^2 R_{rr}(0) + 2 \rho_{21}\rho_{22} R_{rr'}(0) + \rho_{22}R_{r'r'}(0)}
\end{align}
%As can be seen, the computational expense of transforming Equation~\ref{eq:m1} to Equation~\ref{eq:m14} is the computation of the terms in Equation~\ref{eq:m16}, which involves some additions and divisions only. 
\section{Evaluation}
\label{sec:evaluation}

%For a continuous function, $r(t)$, $r'(t)$ describes an infinitesimal rate of change of $r(t)$ in the neighbourhood of $t$. For a discrete function, $r[n]$, the  equivalence of $r'(t)$ is the difference quotient and its resolution in describing the rate of change of $r[n]$ depends on the rate at which the original function is sampled to establish $r[n]$. 

A close examination of Fig.~\ref{fig:acf} reveals that the autocorrelation of the received power is at its peak when $\tau = 0$. This results in $R'_{rr}(0) = 0$, in which case solving Equations~\ref{eq:m3} and \ref{eq:m9} lead to the same simplification. Given $R'_{rr}(0) = 0$, the model parameters in Equation~\ref{eq:m1} can be determined directly from Equations~\ref{eq:m3}:
\begin{align}
    \label{eq:e1}
    \rho_r = \frac{R_{rr}(\tau)}{R_{rr} (0)}    \    \      \rho_{r'} = \frac{R'_{rr}(\tau)}{R_{r'r'} (0)} 
\end{align}
We can employ linear approximation to further simplify Equation~\ref{eq:e1}. By definition:
\begin{equation}
    \label{eq:e2}
   R''_{rr}(0) = \lim_{\tau \to 0} \frac{R'_{rr}(\tau) - R'_{rr} (0)}{\tau}
\end{equation}
In other words, for a small $\tau$, 
\begin{equation}
\label{eq:e3}
    R'_{rr}(\tau) \approx R''_{rr}(0)\tau
\end{equation} 
(since $R'_{rr}(0) = 0$). Because $R_{r'r'}(0) = R''_{rr}(0)$, inserting Equation~\ref{eq:e3} into Equation~\ref{eq:e1} yields:
\begin{equation}
    \label{eq:e4}
    \rho_{r'} = \tau
\end{equation}
Similarly, for a small $\tau$, $R_{rr}(\tau) \approx R_{rr}(0)$, so that $\rho_r = 1$. With the model parameters so determined, the MS estimation of the received power becomes:
\begin{equation}
    \label{eq:e5}
    \hat{\mathbf{r}}(t + \tau) = \mathbf{r}(t)  + \tau \mathbf{r}'(\tau) 
\end{equation}

\subsection{CC2538 System-on-Chip}
\begin{figure}
	\centering
	\includegraphics[ angle = 270, width=0.5\textwidth]{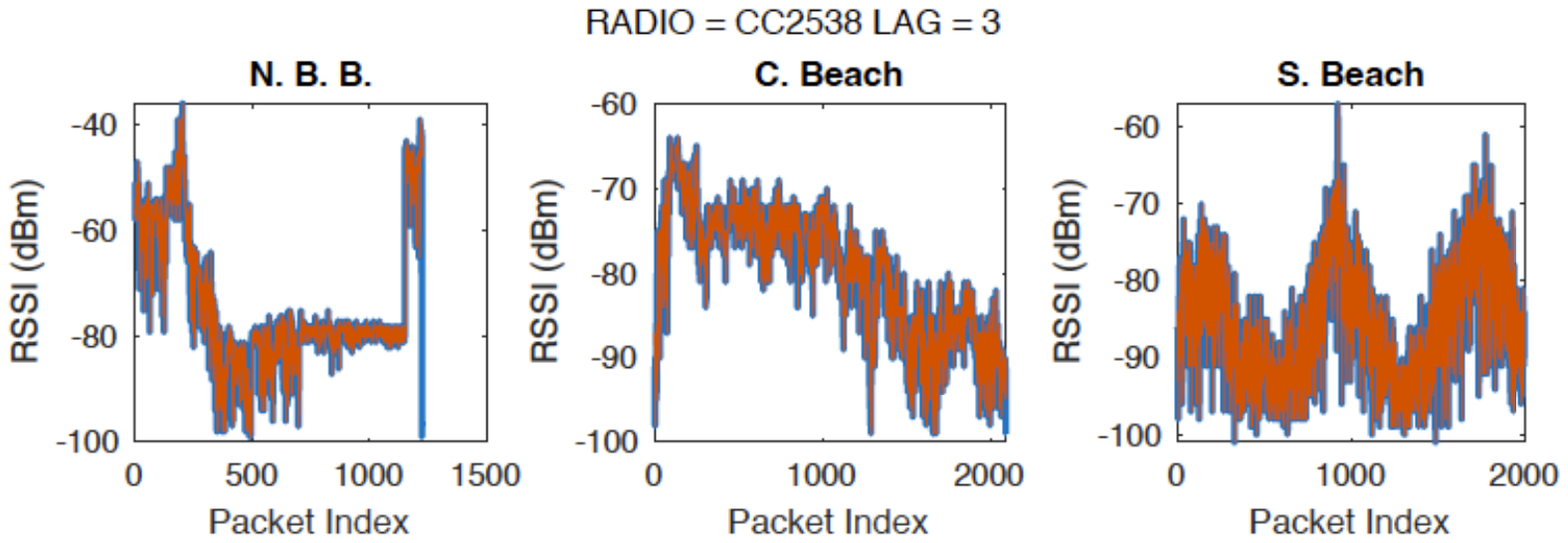}
        \caption{Prediction of the received power for three different deployment environments using the CC2538 SoC. LEFT: North Biscayne Bay. MIDDLE: Miami Crandon Beach. RIGHT: Miami South Beach. LAG = 3.}
	\label{fig:estimation_cc2538}
\end{figure}

\begin{figure}
	\centering
	\includegraphics[ angle = 270, width=0.45\textwidth]{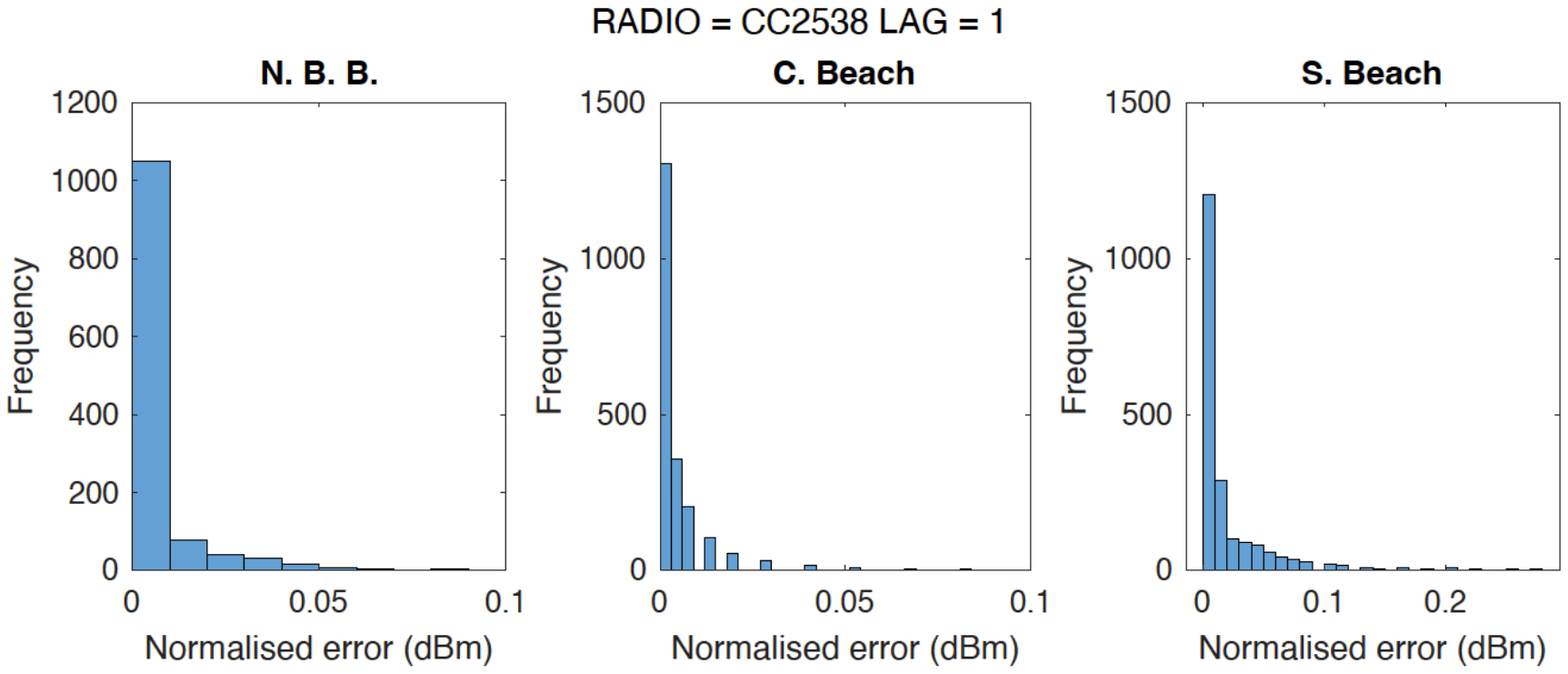}
	\caption{The normalised (between 0 and 1) prediction error of our model for the deployments of Fig.~\ref{fig:estimation_cc2538}.}
	\label{fig:error_cc2538}
\end{figure}

\begin{figure}
	\centering
     \includegraphics[ angle = 270, width=0.45\textwidth]{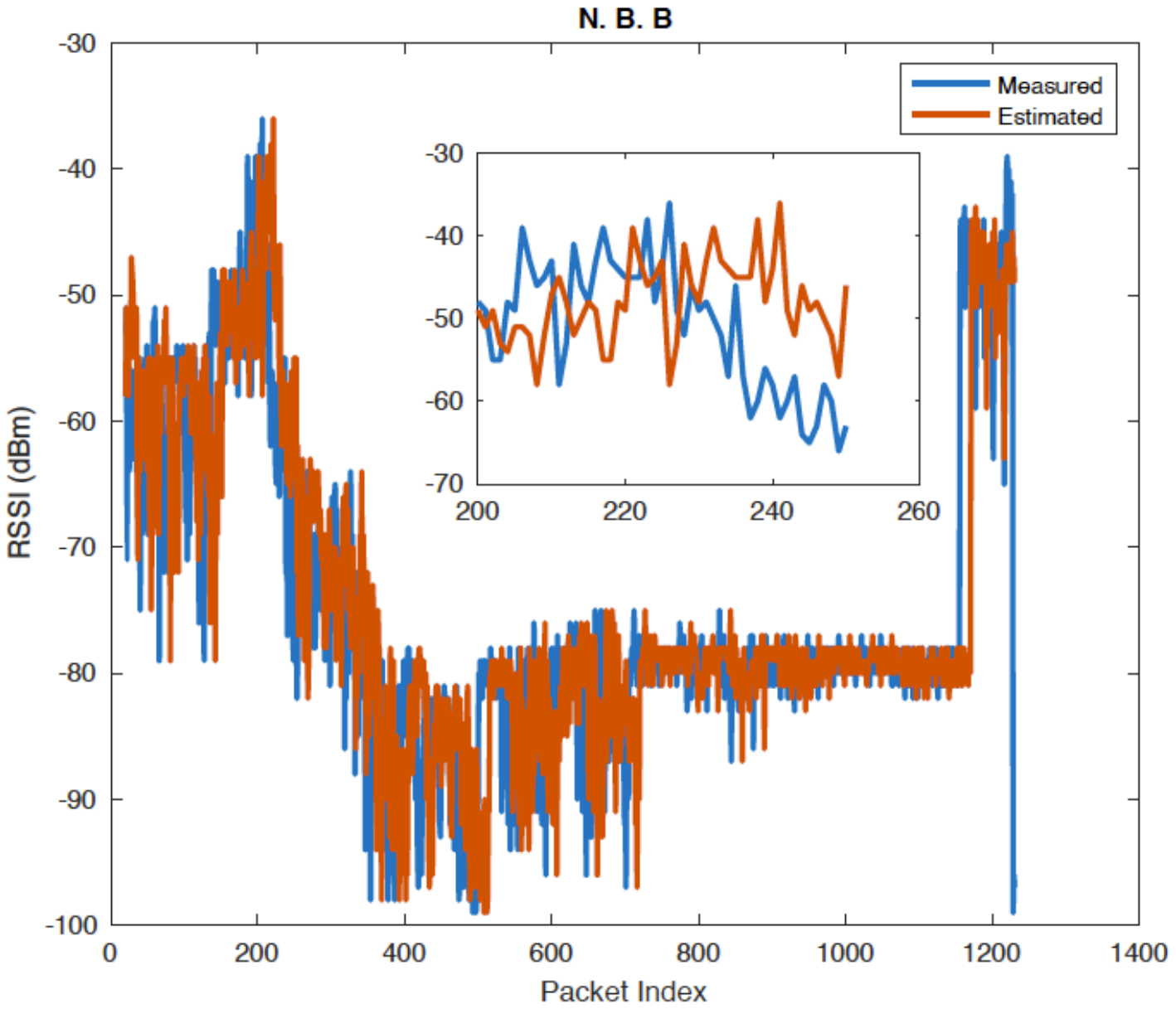}
        \caption{Prediction of the received power for a deployment carried out at North Biscayne Bay using the CC2538 SoC. LAG = 15.}
        \label{fig:estimation_cc2538_lag_15}
\end{figure}

Fig.~\ref{fig:estimation_cc2538} shows the plots of the model's prediction of the received power (in terms of RSSI) for the CC2538 SoC for three of our deployment environments. The plot corresponds to a lag of 3 unit time (since we transferred 10 packets per second with the CC2538 radio, LAG = 1 corresponds to 100 ms and LAG = 3 corresponds to 300 ms). The histograms of the mean square error are plotted in Fig.~\ref{fig:error_cc2538}. The autocorrelation declines when the lag between the present and the future received power increases, as can be seen in Fig.~\ref{fig:acf}. Table~\ref{tab:error_cc2538} summarises the Root Mean Square Error (RMSE) of the model for the different deployment environments and three different lags. Except for Miami South Beach (whose water surface experienced a robust 3D motion), the prediction accuracy is above 90\%. Indeed, based on all the experiments we conducted (five for each of the deployment environment) and lags ($LAG \leq 3$), the average prediction accuracy is 90\%. When the lag becomes too long, the assumptions leading to Equation~\ref{eq:e5} no longer hold and Equation~\ref{eq:m9} should be used instead of Equation~\ref{eq:e5}; even then, the model's prediction error becomes significant, as $R_{rr}(\tau)$ approaches zero. To illustrate this, we plot the model's prediction of the received power for $LAG = 15$ (corresponding to 1.5 s) for one of our deployments (North Biscayne Bay) in Fig.~\ref{fig:estimation_cc2538_lag_15}. The plots are zoomed-in to highlight the estimation inaccuracies. The corresponding RMSE is given in Table~\ref{tab:error_cc2538}.

\begin{table}[t!]
    \caption{The RMSE of the model for the different deployments. Radio: CC2538 SoC. A LAG is 100 $ms$}
    \centering
    \begin{tabular}{|l|c|c|c|c|}
    \hline
    \textbf{Deployment} &     \textbf{LAG = 1} &     \textbf{LAG = 2} & \textbf{LAG = 3} &  \textbf{LAG = 15} \\ 
     \hline
    \hline
    \textbf{N. B. Bay}     & 7.92\%    & 8.77\%    & 9.47\%        & 12.74\%       \\
    \textbf{C. Beach}          & 6.48\%    & 9.21\%    & 10.41\%       & 12.82\%         \\
    \textbf{S. Beach}            & 13.63\%    & 14.75\%    & 15.12\%     & 15.48\%      \\
    \hline 
    \end{tabular}
    \label{tab:error_cc2538}
\end{table}
 \newpage
\subsection{CC1200 Radio Chip}
\begin{figure}
	\centering
		\includegraphics[width=0.45\textwidth]{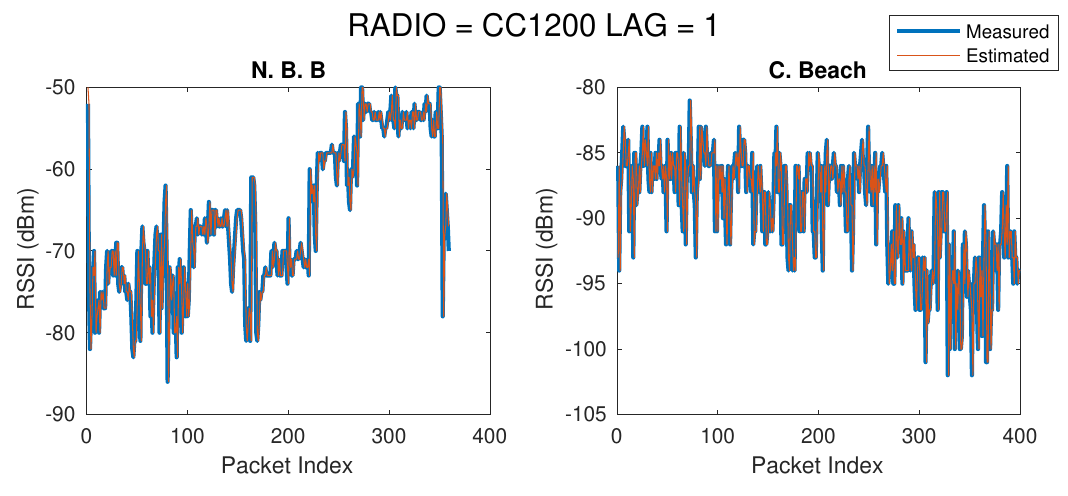}
        \includegraphics[width=0.45\textwidth]{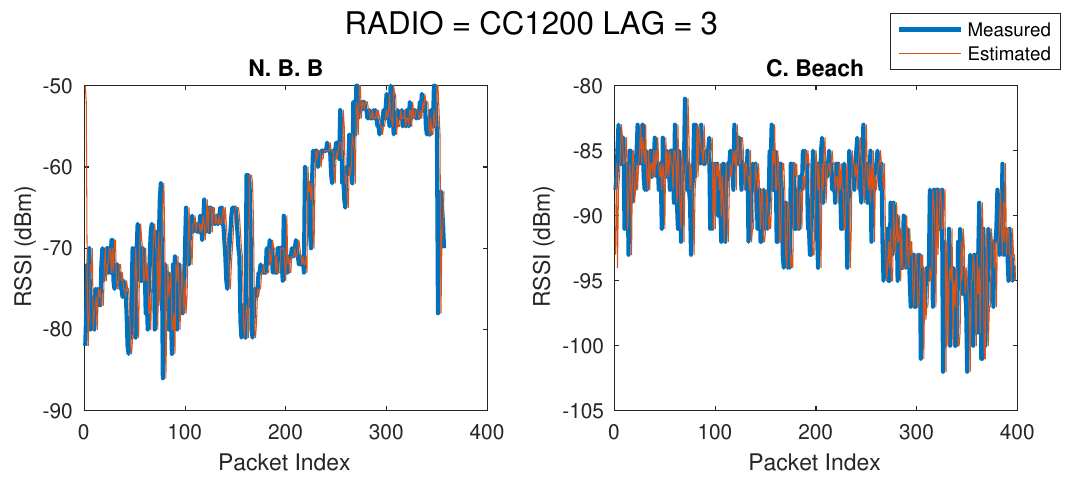}
        \caption{Prediction of the received power for two different deployment environments using the CC1200 radio. Top: LAG = 1. BOTTOM: LAG = 3. LEFT: North Biscayne Bay. RIGHT: Miami Crandon Beach}
	\label{fig:estimation_cc1200}
\end{figure}

\begin{figure}
	\centering
	\includegraphics[width=0.45\textwidth]{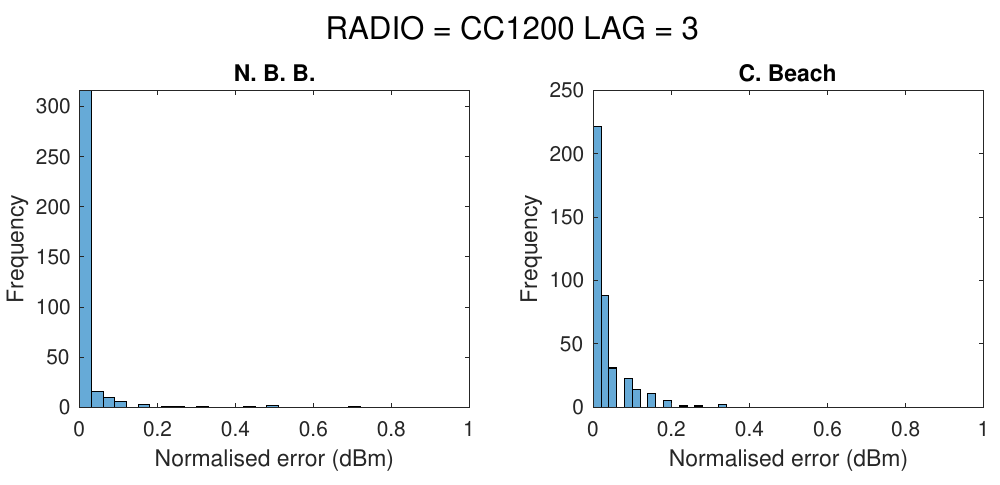}
	\caption{The normalised (between 0 and 1) prediction error of our model for the deployments of Fig.~\ref{fig:estimation_cc1200}.}
	\label{fig:error_cc1200}
\end{figure}

Similarly, Fig.~\ref{fig:estimation_cc1200} displays the plots of the model's prediction of the received power for the CC1200 radio for two of our deployment environments. The plots correspond to LAG = 1 (with a transmission rate of 2 packets per second, this lag corresponds to 500 ms) and LAG = 3 (1.5 s). The histograms of the mean square error for LAG = 1 are plotted in Fig.~\ref{fig:error_cc1200}. Compared to the CC2538, the model's accuracy is slightly degraded (the average prediction accuracy was 85\%). The reason for this is the relatively lower packet transmission rate the radio was able to support (2 Hz compared to the 10 Hz the CC2538 sustainably supported). The impact of low packet transmission rate is that, due to a large time interval between any two packets, the autocorrelation of the received power rapidly falls even for small lags. This problem is further exacerbated by larger lags, as can be seen in Table~\ref{tab:error_cc1200}.  

\begin{table}[t!]
  \caption{The RMSE of the model for the different deployments. Radio: CC1200. A LAG is 500 $ms$}
   \centering
    \begin{tabular}{|l|c|c|c|}
    \hline
    \textbf{Deployment} &     \textbf{LAG = 1} &     \textbf{LAG = 2} & \textbf{LAG = 3} \\ 
     \hline
    \hline
    \textbf{North Biscayne Bay}     & 15.45\%  & 16.60\%    & 17.52\%    \\
    \textbf{Crandon Beach}          & 10.56\%    & 13.87\%    & 15.65\% \\
    \hline 
    \end{tabular}
    \label{tab:error_cc1200}
\end{table}

%For the evaluation of our model, we used the traces of the received power (RSSI values) for all the deployments we carried out in the four different water bodies and the two different radios. 
\subsection{Comparison}

We compare the performance of our model with the performance of the models proposed in \cite{cao2020optimized} and \cite{lin2016atpc}. In the first, the mean square error ranges from a value slightly less that $10^{-1}$ to a value approaching $10^{-3}$, depending on the desired trade-off between model complexity, signal-to-noise ratio, and the signal denoising (reconstruction) factor. By contrast, the prediction accuracy we achieved is smaller, but to compensate for that, our model does not make any assumption about the topology of the underlying network and the wireless channel. Secondly, our results are based on practical deployments, achieved after overcoming several practical challenges, which are not articulated (and directly addressed) in \cite{cao2020optimized}. Similarly, the model in \cite{lin2016atpc} achieves an impressive performance. Even though the authors do not offer a specific prediction accuracy, it is reported that the model reduces power consumption by up to  53.6\%  and achieves a Packet Reception Ratio of 99\%. However, the proposed model deals with static deployments wherein the transmission power is affected by factors which change slowly over time (such as shadowing and people passing). For our case, the deployment environments are considerably harsher. Moreover, the model is a so-called ``one-step ahead'' and can be computationally intensive when the parameter to be predicted changes slowly over time, as it makes prediction for each future sample. So called ``n-step ahead'' models, by contrast, can be configured to look n-step ahead. In other words, if the link quality is good, frequent prediction is not necessary and the prediction interval can be adjusted by choosing the appropriate lag. In \cite{Dargie10713111}, a Kalman Filter is proposed to estimate received power. The model requires measurement and process error statistics, both of which are established at a modest cost. The measurement error was established prior to deployment, using static fingerprinting and the process error statistics were established after transmitting packets in burst for a few minutes. The model, configured as a ``one-step ahead'' predictor for the CC2538 radio, achieved a 90\% accuracy on average. With a similar configuration (LAG = 1 = 100 ms), the present model achieved a comparable result with much less complexity.

%More importantly, both previous models involve matrix inversion whose computation complexity is in the order of $\mathcal{O}(N^{2.376})$, where $N$ is the dimension of the matrix \cite{williams2012multiplying}. The present model avoids matrix inversion in two ways, namely, through (1) model transformation (orthonormalisation) and (2) function approximation.

\section{Conclusion} % 1
\label{sec:conclusion}

In this paper we proposed a lightweight ``n-step'' predictor to estimate the received power of low-power sensing devices deployed in harsh environment. Prediction of the received power is essential to support dynamic transmission power control. The predictor enables to estimate the received power of incoming packets in the presence of successively lost packets. Even though the predictor is based on Minimum Mean Square Estimation, it avoids matrix inversion to determined the model parameters through two essential estimation steps, namely, function approximation and orthonormalisation. Based on practical deployments of low-power sensing nodes we carried out on four different water bodies and using two different types of low-power radios, we demonstrated that the model achieved a prediction accuracy exceeding 90\%. A further improvement of the prediction accuracy is possible, but this comes with an increased computational cost. Our future research focus is to closely investigate this possibility.

\balance
\bibliographystyle{IEEEtran}
\bibliography{library}

\begin{thebibliography}{10}
\providecommand{\url}[1]{#1}
\csname url@samestyle\endcsname
\providecommand{\newblock}{\relax}
\providecommand{\bibinfo}[2]{#2}
\providecommand{\BIBentrySTDinterwordspacing}{\spaceskip=0pt\relax}
\providecommand{\BIBentryALTinterwordstretchfactor}{4}
\providecommand{\BIBentryALTinterwordspacing}{\spaceskip=\fontdimen2\font plus
\BIBentryALTinterwordstretchfactor\fontdimen3\font minus
  \fontdimen4\font\relax}
\providecommand{\BIBforeignlanguage}[2]{{%
\expandafter\ifx\csname l@#1\endcsname\relax
\typeout{** WARNING: IEEEtran.bst: No hyphenation pattern has been}%
\typeout{** loaded for the language `#1'. Using the pattern for}%
\typeout{** the default language instead.}%
\else
\language=\csname l@#1\endcsname
\fi
#2}}
\providecommand{\BIBdecl}{\relax}
\BIBdecl

\bibitem{dargie2010fundamentals}
W.~Dargie and C.~Poellabauer, \emph{Fundamentals of wireless sensor networks:
  theory and practice}.\hskip 1em plus 0.5em minus 0.4em\relax John Wiley \&
  Sons, 2010.

\bibitem{li2020electromagnetic}
Y.~Li, J.~Wu, and Q.~Guo, ``Electromagnetic sensor for detecting wear debris in
  lubricating oil,'' \emph{IEEE Transactions on Instrumentation and
  Measurement}, vol.~69, no.~5, pp. 2533--2541, 2020.

\bibitem{wang2020active}
X.~Wang, ``Active fault tolerant control for unmanned underwater vehicle with
  sensor faults,'' \emph{IEEE Transactions on Instrumentation and Measurement},
  vol.~69, no.~12, pp. 9485--9495, 2020.

\bibitem{Trihinas8057144}
D.~Trihinas, G.~Pallis, and M.~D. Dikaiakos, ``Admin: Adaptive monitoring
  dissemination for the internet of things,'' in \emph{IEEE INFOCOM 2017 - IEEE
  Conference on Computer Communications}, 2017, pp. 1--9.

\bibitem{azarhava2020energy}
H.~Azarhava and J.~M. Niya, ``Energy efficient resource allocation in wireless
  energy harvesting sensor networks,'' \emph{IEEE Wireless Communications
  Letters}, vol.~9, no.~7, pp. 1000--1003, 2020.

\bibitem{dargie2011dynamic}
W.~Dargie, ``Dynamic power management in wireless sensor networks:
  State-of-the-art,'' \emph{IEEE Sensors Journal}, vol.~12, no.~5, pp.
  1518--1528, 2011.

\bibitem{Ju7842599}
Q.~Ju and Y.~Zhang, ``Predictive power management for internet of battery-less
  things,'' \emph{IEEE Transactions on Power Electronics}, vol.~33, no.~1, pp.
  299--312, 2018.

\bibitem{liu2020multistep}
F.~Liu, C.~Jiang, and W.~Xiao, ``Multistep prediction-based adaptive dynamic
  programming sensor scheduling approach for collaborative target tracking in
  energy harvesting wireless sensor networks,'' \emph{IEEE Transactions on
  Automation Science and Engineering}, vol.~18, no.~2, pp. 693--704, 2020.

\bibitem{Arghavani10404000}
A.~Arghavani, H.~Zhang, Z.~Huang, and Y.~Chen, ``Power-adaptive communication
  with channel-aware transmission scheduling in wbans,'' \emph{IEEE Internet of
  Things Journal}, vol.~11, no.~9, pp. 16\,087--16\,102, 2024.

\bibitem{Zhao10577633}
C.~Zhao, B.~Tang, and L.~Deng, ``Missing-measurements-tolerant compressed
  sensing in wireless sensor networks for mechanical vibration monitoring,''
  \emph{IEEE Transactions on Instrumentation and Measurement}, vol.~73, pp.
  1--13, 2024.

\bibitem{clerckx2021wireless}
B.~Clerckx, K.~Huang, L.~R. Varshney, S.~Ulukus, and M.-S. Alouini, ``Wireless
  power transfer for future networks: Signal processing, machine learning,
  computing, and sensing,'' \emph{IEEE Journal of Selected Topics in Signal
  Processing}, vol.~15, no.~5, pp. 1060--1094, 2021.

\bibitem{shahraki2020clustering}
A.~Shahraki, A.~Taherkordi, {\O}.~Haugen, and F.~Eliassen, ``Clustering
  objectives in wireless sensor networks: A survey and research direction
  analysis,'' \emph{Computer Networks}, vol. 180, p. 107376, 2020.

\bibitem{amutha2020wsn}
J.~Amutha, S.~Sharma, and J.~Nagar, ``Wsn strategies based on sensors,
  deployment, sensing models, coverage and energy efficiency: Review,
  approaches and open issues,'' \emph{Wireless Personal Communications}, vol.
  111, no.~2, pp. 1089--1115, 2020.

\bibitem{dargie2024estimation}
W.~Dargie, ``Estimation of motion statistics from statistics of received power
  in low-power iot sensing nodes,'' \emph{IEEE Sensors Letters}, 2024.

\bibitem{hu2020tngan}
X.~Hu, H.~Zhang, D.~Ma, and R.~Wang, ``A tngan-based leak detection method for
  pipeline network considering incomplete sensor data,'' \emph{IEEE
  Transactions on Instrumentation and Measurement}, vol.~70, pp. 1--10, 2020.

\bibitem{zonzini2020vibration}
F.~Zonzini, M.~M. Malatesta, D.~Bogomolov, N.~Testoni, A.~Marzani, and
  L.~De~Marchi, ``Vibration-based shm with upscalable and low-cost sensor
  networks,'' \emph{IEEE Transactions on Instrumentation and Measurement},
  vol.~69, no.~10, pp. 7990--7998, 2020.

\bibitem{park2020adaptable}
E.~Park, M.-S. Lee, H.-S. Kim, and S.~Bahk, ``Adaptable: Adaptive control of
  data rate, transmission power, and connection interval in bluetooth low
  energy,'' \emph{Computer Networks}, vol. 181, p. 107520, 2020.

\bibitem{cao2020optimized}
X.~Cao, G.~Zhu, J.~Xu, and K.~Huang, ``Optimized power control for over-the-air
  computation in fading channels,'' \emph{IEEE Transactions on Wireless
  Communications}, vol.~19, no.~11, pp. 7498--7513, 2020.

\bibitem{zang2017gait}
W.~Zang and Y.~Li, ``Gait-cycle-driven transmission power control scheme for a
  wireless body area network,'' \emph{IEEE journal of biomedical and health
  informatics}, vol.~22, no.~3, pp. 697--706, 2017.

\bibitem{lin2016atpc}
S.~Lin, F.~Miao, J.~Zhang, G.~Zhou, L.~Gu, T.~He, J.~A. Stankovic, S.~Son, and
  G.~J. Pappas, ``Atpc: Adaptive transmission power control for wireless sensor
  networks,'' \emph{ACM Transactions on Sensor Networks (TOSN)}, vol.~12,
  no.~1, pp. 1--31, 2016.

\bibitem{zhang2022modern}
X.-D. Zhang, \emph{Modern signal processing}.\hskip 1em plus 0.5em minus
  0.4em\relax Walter de Gruyter GmbH \& Co KG, 2022.

\bibitem{papoulis2002probability}
A.~Papoulis and S.~Unnikrishna~Pillai, \emph{Probability, random variables and
  stochastic processes}.\hskip 1em plus 0.5em minus 0.4em\relax McGraw-Hill
  Higher Education (4. edition), 2002.

\bibitem{leon2013gram}
S.~J. Leon, {\AA}.~Bj{\"o}rck, and W.~Gander, ``Gram-schmidt orthogonalization:
  100 years and more,'' \emph{Numerical Linear Algebra with Applications},
  vol.~20, no.~3, pp. 492--532, 2013.

\bibitem{Dargie10713111}
W.~Dargie, P.~Padrao, L.~Bobadilla, and C.~Poellabauer, ``Link quality
  fluctuation in wireless networks deployed on the surface of different water
  bodies,'' \emph{IEEE Sensors Journal}, vol.~24, no.~23, pp. 39\,789--39\,797,
  2024.

\end{thebibliography}

\end{document}